\documentclass[prd,showpacs,amsmath,showkeys,twocolumn,floatfix,amssymb, preprintnumbers, nofootinbib, superscriptaddress]{revtex4} 
\usepackage{epsfig,dcolumn}
\usepackage{graphicx}
\usepackage{comment} 
\usepackage{verbatim}
\usepackage[usenames]{color}
\usepackage{graphicx}
\usepackage{bm}
 \usepackage{ifpdf}
 \usepackage{multirow}
\usepackage{soul}

\def\lsim{\mathrel{\rlap{\lower4pt\hbox{\hskip1pt$\sim$}}
    \raise1pt\hbox{$<$}}}
\def\gsim{\mathrel{\rlap{\lower4pt\hbox{\hskip1pt$\sim$}}
    \raise1pt\hbox{$>$}}}
\begin{document}

\title{Dispersive approaches for three-particle  final states interaction}

\author{Peng~Guo}
\email{pguo@jlab.org}
\affiliation{Physics Department, Indiana University, Bloomington, IN 47405, USA}
\affiliation{Center For Exploration  of Energy and Matter, Indiana University, Bloomington, IN 47408, USA.}
\affiliation{Thomas Jefferson National Accelerator Facility, 
Newport News, VA 23606, USA}
\author{I.~V.~Danilkin}
\affiliation{Thomas Jefferson National Accelerator Facility,  
Newport News, VA 23606, USA}
\author{Adam~P.~Szczepaniak}
\affiliation{Physics Department, Indiana University, Bloomington, IN 47405, USA}
\affiliation{Center For Exploration  of Energy and Matter, Indiana University, Bloomington, IN 47408, USA.}
\affiliation{Thomas Jefferson National Accelerator Facility,  
Newport News, VA 23606, USA}

\preprint{JLAB-THY-14-1953}

\date{\today}

\begin{abstract} 
In this work, we present different representations of Khuri-Treiman equation and discuss advantages and disadvantages of each representation.  In particular we focus on the inversion technique proposed by Pasquier, which even though developed a long time ago, has not been used in modern analyses of data on three particle decays.  We apply the method to a toy model and compare the sensitivity of this and alternative solution methods to  the left-hand cut contribution.
We also discuss the meaning and applicability of the Watson's theorem when three particles in final states are involved.  
\end{abstract}

\pacs{ }

\maketitle


\section{Introduction}
\label{intro} 
 
Low energy hadronic reactions play an important role in constraining parameters of effective theories of QCD. Furthermore precision in the determination of the underlying reaction amplitudes is needed when searching for physics beyond the Standard Model.  There are model independent restrictions that at low energies help to determine hadronic amplitudes. These include, for example, chiral symmetry, partial wave and effective range expansions and unitarity. These constraints, together with the requirement that reaction amplitudes respect crossing relations and are analytical functions of the kinematical variables are often used to formulate dispersion relations for the amplitudes. A particular implementation  leads to the so-called Khuri-Treiman (KT) equations \cite{Khuri:1960kt,Bronzan:1963kt,Aitchison:1965kt,Aitchison:1965dt,Aitchison:1966kt,Pasquier:1968kt,Pasquier:1969kt}. These equations were originally written for amplitudes involving four external particles, {\it i.e} amplitudes describing \mbox{$2\to2$} scattering or \mbox{$1\to 3$} decays. Specifically, the KT equations impose two-body unitarity on a truncated set of partial waves in the three channels of a four-point amplitude. In practical applications, for example to the analysis of \mbox{$\eta\to 3\pi$} decay, the KT approach helped reduce the discrepancy between the measured decay width and the NLO chiral perturbation theory  prediction \cite{Kambor:1996,Anisovich:1996,Lanz:2013ku,Kampf:2011wr,Descotes-Genon:2014tla,Bijnens:2002,Danilkin:2014cra,Schneider:2011}. With the availability of high-precision data on production and decays of light hadrons, there has recently been a renewed interest in applications of the KT framework.

Different representations of KT equations and various methods for solving KT equations have been proposed in the past \cite{Khuri:1960kt,Bronzan:1963kt,Aitchison:1965kt,Aitchison:1965dt,Aitchison:1966kt,Pasquier:1968kt,Pasquier:1969kt}.  To the best of our knowledge, the comparison of those different variants of KT equations has never been demonstrated before. In this paper, we begin with the discussion of the various representations and methods of obtaining their solutions.  In particular, we focus on the Pasquier inversion technique that was introduced in \cite{Pasquier:1968kt,Pasquier:1969kt}.  
The solution of the KT equation requires  knowledge of two-body partial wave amplitudes.
Once these amplitudes are known, different methods for solving the equations yield the same answer. 
However, when the Pasquier inversion is applied the input two-body amplitude  
 is probed in a different energy region than in the case of a direct solution of the KT equation. 
 Therefore, for physical applications (where the two-body partial wave amplitudes are only known in a limited energy range), various solution strategy are complimentary to each other and allow to study systematic uncertainties.

 Since the two-body partial wave amplitudes are only known in a limited energy range, in the case of the Pasquier inversion method approximations are often needed when dealing with the left-hand cuts. In \cite{Guo:2015zqa} the authors showed that based on a realistic set of $\pi\pi$ partial waves and a single real parameter representing the left-hand cut, one could effectively reproduce the Dalitz plot distribution in \mbox{$\eta\rightarrow \pi^+\pi^-\pi^-$}. Given the number of unknown parameters in this approximation, it could be beneficial to study the decays of higher mass particles, {\it e.g.} $\eta'$, $\omega$, $\phi$, $D$ {\it etc.} where there are no specific low energy constraints. However, from the other side it is important to illustrate the Pasquier inversion technique and the corresponding approximations in a schematic model where the exact solution is known. This will shed further light on its usefulness.

We consider a toy model of scalar particles interacting in a single partial wave. There are known cases, {\it e.g.}  \mbox{$\omega/\phi\rightarrow 3\pi$} where a single partial wave dominates.  Also in our toy model, the left-hand cut is chosen to be close to the physical region. That is similar to what happens with the $\pi\pi$ amplitudes, where the left-hand starts at $s=0$ and the threshold is located at \mbox{$4\,m_\pi^2$}.

The paper is organized as follows. In Section \ref{KTmodel} we first introduce kinematics and Khuri-Treiman (KT) equations needed for a dispersion description of the three-body decay. Then we discuss different solution strategies and compare numerical results on the hand of the schematic model. As a further analysis, we show how the solutions from various representations of the KT equations change depending on the left-hand cut approximations. In Section \ref{watson} we give a brief discussion of the role of the Watson's final state  interaction theorem.  Summary and conclusion are presented in Section \ref{summary}.

\begin{figure}
\includegraphics[width=0.48\textwidth]{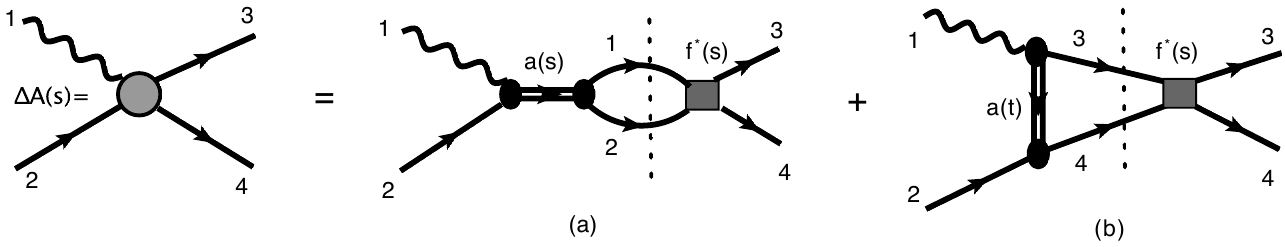}
\caption{A diagrammatic representation of discontinuity relations in Eq.(\ref{dis1}). 
\label{fig:dis}}
\end{figure}

\section{The Khuri-Treiman Model}\label{KTmodel} 
\subsection{Definition} 

\begin{figure}
\includegraphics[width=0.48\textwidth]{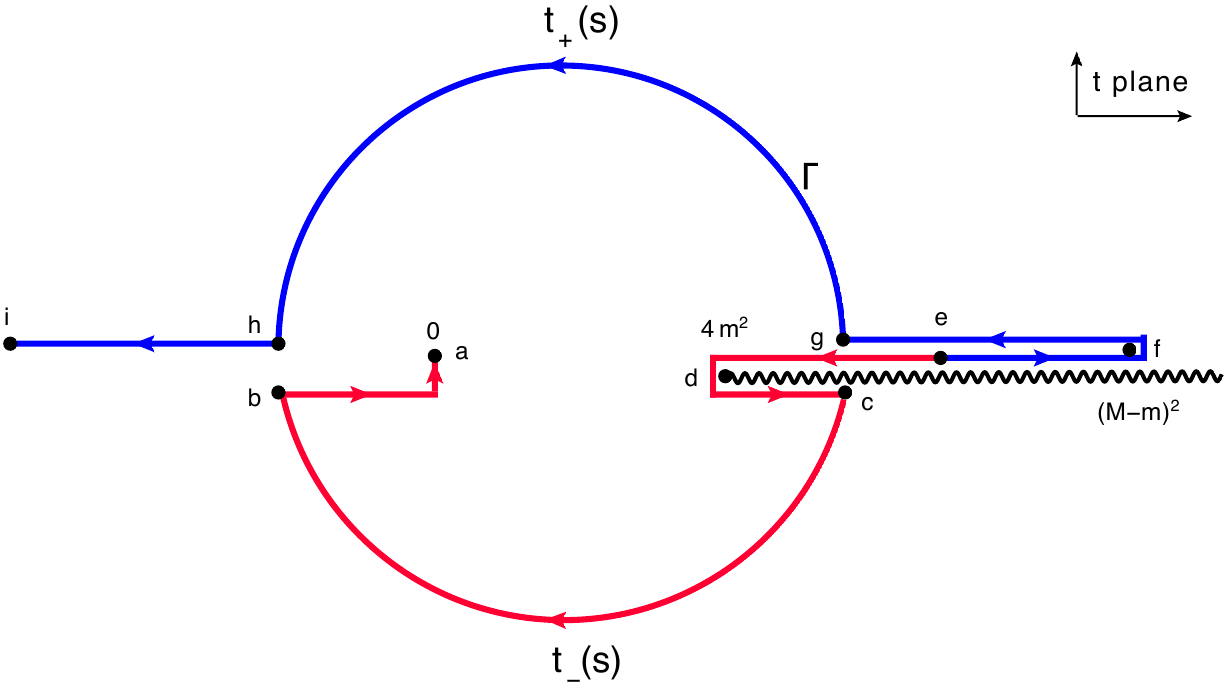}
\caption{  Motion of the upper, $t_+(s)$ and lower, $t_-(s)$ limits of integration in Eq.(\ref{dis1}) in the complex plane as a function of (real) $s$. The arrows indicate the direction of increasing $s$  in the interval from \mbox{$4\,m^{2}$} to $\infty$.  The zigzag line represents the location of singularities (right hand cut) of  $a(t)$. The  points  labeled 
 $a$ through $i$  correspond to specific values of $s$, with  
 (a) $t_{-}( \infty)= 0$, (b) $t_{-}( (M+m)^{2})=  m\,(m-M)$,  (c) $t_{-}((M-m)^{2} )=  m\,(M+m)$, (d) $t_{-}( \frac{ M^{2} -m^{2} }{2})=4\,m^{2}$, (e) $t_{\pm}( 4\,m^{2})= \frac{ M^{2} -m^{2} }{2}$, (f) $t_{+}( m\,(M+m))= (M-m)^{2}$,  (g) $t_{+}( (M-m)^{2})=  m\,(M+m)$, (h) $t_{+}((M+m)^{2})=  m\,(m-M)$,  and (i) $t_{+}(\infty)= -\infty$, respectively.
\label{fig:2}}
\end{figure}

The reaction amplitude describing interactions of four scalar particles \mbox{$i=1 \cdots 4$} is given by a single complex function, $A(s,t,u)$, of the three Mandelstam variables  satisfying \mbox{$s+t+u = M^2 + 3m^2$}. In the following  $i=2,3,4$ labels identical particles with mass $m$ and \mbox{$i=1$} refers to a particle with mass $M$. We are specifically interested in the decay of particle $1$ {\it i.e.} the case \mbox{$M>3\,m$}. The  amplitude $A(s,t,u)$ describes four distinct processes. These are (with the bar denoting an antiparticle) the $s$-channel scattering, \mbox{$1 + 2 \to 3 + 4$}, the $t$-channel scattering \mbox{$1 + \bar 3 \to \bar 2 + 4$}, the $u$-channel scattering, \mbox{$1 + \bar 4 \to 3 + \bar 2$} and the particle $1$ decay channel, \mbox{$1\to \bar 2 + 3  + 4$}.  In terms of particle momenta the three Mandelstam variables are \mbox{$s=(p_1 + p_2)^2$}, \mbox{$t=(p_1 + \bar p_3)^2 = (p_1 - p_3)^2$}, \mbox{$u = (p_1 + \bar p_4)^2 = (p_1 - p_4)^2$}. In the $s$-channel, the partial wave (p.w.) expansion,   
\begin{equation} 
A(s,t,u) = 16\pi \sum_{l=0}^\infty  (2l+1)\,A_l(s)\,P_l(z_s),  \label{1} 
\end{equation} 
defines the $s$-channel partial waves, $A_l(s)$,  with \mbox{$z_s = \cos\theta_s$} being the cosine of the $s$-channel center of mass scattering angle, 
\begin{equation} 
z_s =  \frac{s\,(t - u)}{ \lambda^{1/2}(s,M^2,m^2) \lambda^{1/2}(s,m^2,m^2) }\,. \label{zs} 
\end{equation} 
The triangle function $\lambda$ is  given by \mbox{$\lambda(a,b,c) = a^2 + b^2 + c^2 - 2\,a\,b - 2\,b\,c - 2\,a\,c$}. Similar p.w. expansions can be written in the $t$- and the $u$-channel with the center of mass scattering angles given by 
\begin{eqnarray} 
& & z_t  =  \frac{t\,(s - u)}{ \lambda^{1/2}(t,M^2,m^2)\lambda^{1/2}(t,m^2,m^2) }, \nonumber \\
& & z_u =  \frac{u\,(t - s)}{\lambda^{1/2}(u,M^2,m^2) \lambda^{1/2}(u,m^2,m^2)}, 
\end{eqnarray}
respectively. 
A truncated p.w. series defines an amplitude which is a regular function of the scattering angle and its only singularities are with respect to the channel energy variable, {\it i.e.} the $s$-channel p.w. expansion has singularities in $s$ but not in $z_s$ or equivalently $t$ or $u$.   Since $A(s,t,u)$ has singularities in all three variables, it implies that p.w. series in any channel diverges outside the physical region of that channel {\it e.g.} the $s$-channel p.w. expansion diverges outside the \mbox{$s>(M+m)^2>4m^2$} and \mbox{$|z_s| < 1$}. In other words, p.w. expansion in a specific channel needs to be analytically continued outside the physical region where the series is defined to obtain the physical amplitude representing reactions in the other channels. In the Khuri-Treiman model singularities  of $A(s,t,u)$ in all three variables are recovered by approximating the amplitude as the sum of p.w. series in the three channels simultaneously \cite{Khuri:1960kt,Bronzan:1963kt,Aitchison:1965kt,Aitchison:1965dt,Aitchison:1966kt,Pasquier:1968kt,Pasquier:1969kt}, with each sum truncated at some finite value, $l=L$, 
\begin{align}
A(s,t,u) & =   16\,\pi \sum_{l=0}^{L} (2l+1)    \nonumber \\
& \times   \left[ a_l(s)\,P_l(z_s)  + a_l(t)\,P_l(z_t) + a_l(u)\,P_l(z_u) \right].   \label{KT}
\end{align}
The truncation on the number of partial waves alleviates the problem of double counting, since the removed, high partial waves in one channel are being replaced by the low partial waves in the  crossed channels. In Eq.~(\ref{KT}), $a_l$'s denote the partial wave amplitudes of the KT model and should be distinguished from the $A_l$'s in Eq.(\ref{1}). The relation between the two is obtained by projecting Eq.~(\ref{KT}) on to the $s$-channel partial waves, 
\begin{align} 
    A_l(s) = a_l(s) +& \frac{1}{2} \int_{-1}^{1}  dz_s  P_l(z_s) \sum_{l'=0}^L (2 l' + 1)\nonumber \\
   &\times  \left[ a_{l'}(t(s,z_s))\,P_{l'}(z_t(s,z_s)) 
   + (t \leftrightarrow u) \right]. 
\end{align} 
Each of the three terms on the right hand side of Eq.(\ref{KT}) has singularities in one variable only {\it i.e.}  $s$, $t$ and $u$, respectively. These singularities are assumed to originate from unitarity, so that even though  $a_l(s)$ has only the right hand, unitary cut, $A_l(s)$ also has the left hand cut due to the exchange terms. Truncation of the partial wave series  leads to an incorrect asymptotic behavior of $A(s,t,u)$ at large values of the Mandelstam variables and the KT model is intrinsically limited to low-energies. For simplicity, in the following analysis we truncate the partial waves to include $S$-waves only {\it i.e.} set \mbox{$L = 0$} and denote \mbox{$a(s) \equiv a_{0} (s)$}. At low energies, below the first inelastic threshold, unitarity is saturated by two-particle intermediate states, the discontinuity relation of $a(s)$ by crossing unitarity cut in $s$ is thus given by,
\begin{align} 
\Delta a(s) & = \frac{1}{2i}\left( a(s+ i\epsilon) - a(s - i\epsilon)\right)  \nonumber \\
&  = f^*(s)\,\rho(s) \left[ a(s) +\frac{2}{K(s)} \int_{t_-(s)}^{t_+(s)} dt \, a(t)   \right].  \label{dis1} 
\end{align}
Here \mbox{$\rho(s) = \sqrt{1 - 4\,m^2/s}$} is the two body phase space factor and $f(s)$ is the $S$-wave scattering amplitude describing two-body interactions between pairs of the particles $1,2$ and $3$.  The first term on the right hand side of Eq.(\ref{dis1}) corresponds to the case when particles \mbox{$3+4$} are produced from  particles \mbox{$1+2$} in $s$-channel, $S$-wave projection of the unitarity relation. It is illustrated in the diagram in Fig.~\ref{fig:dis}(a). One half of the second term, illustrated in Fig.~\ref{fig:dis}(b), gives the contribution from the \mbox{$1+2\to 3+4$} by $t$-channel exchange,  and it is obtained by projecting the $t$-channel  two-particle intermediate state expressed through the $t$-channel partial wave series  ({\it i.e.} the second term on {\it r.h.s} of Eq.(\ref{KT}) truncated to include $S$-waves only) onto the $s$-channel $S$-wave.  The factor of $2$ in front of the integral takes into account the  contribution for the $u$-channel (last term on the {\it r.h.s} in Eq.(\ref{KT})).  The $s$-channel partial wave projection of the cross channel partial wave series is obtained by integrating the $t$-channel amplitude over $z_s$, which, using Eq.(\ref{zs}) is expressed as an integral over $t$ at fixed $s$. The integration limits $t_{\pm}(s)$ correspond to $z_s = \pm 1$,
\begin{equation} 
t_{\pm}(s) = \frac{M^2 + 3m^2 -s}{2} \pm \frac{K(s)}{2}\, ,
\end{equation} 
where $K(s)$ is given by \cite{Kambor:1996},
\begin{align}
K(s)=&\left\lbrace
\begin{array}{ll}
+\kappa(s)\,,&\quad  4\,m^2\leq s\leq (M-m)^2 , \\
\,\,i\,\kappa(s)\,,&\quad (M-m)^2\leq s\leq (M+m)^2, \\
-\kappa(s)\,,&\quad (M+m)^2\leq s<+\infty ,
\end{array}
\right.\nonumber \\
\kappa(s)=&\frac{1}{s}\,|\lambda(s,M^2,m^2)\,\lambda(s,m^2,m^2)|^{1/2}\,.
\end{align}
For $s$ real and \mbox{$s>4m^2$}, the integration of $t$-argument in Eq.(\ref{dis1}) follows the path shown in Fig.~\ref{fig:2}. In particular when $s$ is decreased towards \mbox{$s \to (M-m)^2$} {\it i.e.} it approaches the physical boundary of the decay region \mbox{$1 \to \bar 2 + 3 + 4$}, as expected, the integration path followed by $dt$ approaches the positive real axis. Even though, however, at \mbox{$s=(M-m)^2$}, \mbox{$t_+(s) = t_-(s)$}  the integration path remains finite as it runs below the unitary cut, from point $c$ on Fig.~\ref{fig:2} though point $d$ to point $g$ above the $t$-channel unitarity cut of $a(t)$. This was the key observation made in \cite{Bronzan:1963kt} where it was also  shown to be consistent with perturbation theory. That is to say, $a(s)$ in the decay region is obtained by an analytical continuation from the scattering region. This is the case because the integration path shown in Fig.~\ref{fig:2} avoids any singularities as $s$ is decreased from the scattering, \mbox{$s>(M+m)^{2}$} to the domain of the decay region \mbox{$4m^2 < s < (M-m)^2$}.

In summary, the KT model is a low-energy approximation to the amplitude $A(s,t,u)$ in which the exchange forces are approximated by a finite number of partial waves. Given the discontinuity relation Eq.(\ref{dis1}), $a(s)$ can be constructed by using  Cauchy theorem (assuming no subtractions are needed), 
\begin{equation} 
a(s) = \frac{1}{\pi} \int_{4m^2}^\infty ds' \frac{\Delta a(s')}{s' -s} \label{dis2} 
\end{equation}
with input to this equation provided by the two-body   partial wave, $f(s)$.

\subsection{Methods for solving Eq.(\ref{dis2})} 
\label{methods} 
If the exchange term in Eq.(\ref{dis1}) is ignored, the resulting discontinuity relation for $a(s)$ is of the Muskhelishvili-Omn\`es  type \cite{Muskhelishvili,Omne} and can be solved 
using the standard $N/D$ method \cite{Chew:1960nd,Frye:1963nd}. Elastic unitarity determines the discontinuity of the $S$-wave two-body scattering amplitude, 
\begin{equation} 
\Delta f(s) =  \frac{1}{2i} \left( f(s+i\epsilon) - f(s-i\epsilon)\right)  = \rho(s)\,|f(s)|^2. \label{ut} 
\end{equation} 
The algebraic solution of Eq.(\ref{ut}) is 
\begin{equation} 
f(s) = \frac{e^{i\,\delta(s)} \sin\delta(s) }{\rho(s)}, \label{ps}
\end{equation} 
where $\delta(s)$ is the $S$-wave phase shift. Writing the  partial wave amplitude as \mbox{$f(s) = N(s)/D(s)$}, with $D(s)$ containing the elastic cut and all other cuts absorbed into the function $N(s)$, linearizes Eq.(\ref{ut}) and yields an analytical parametrization for $f(s)$,  with 
\begin{equation} 
D(s) = \exp\left(- \frac{s}{\pi}\int_{4m^2}^\infty ds' \frac{\delta(s')}{s'(s'-s)} \right). \label{nd} 
\end{equation} 
By convention we  normalized $D(s)$ so that \mbox{$D(0) = 1$}. The function $1/D(s)$ is referred to as the Muskhelishvili-Omn\`es (MO)  function \cite{Muskhelishvili,Omne}. Consistency with Eqs.(\ref{ut}-\ref{ps}) requires that $N(s)$ is a regular function above elastic threshold. From Eq.(\ref{nd}) it follows that the asymptotic behavior of $D(s)$ for \mbox{$|s| \to \infty$} is given by $s^{\delta(\infty)/\pi}$ .

Writing 
\begin{equation} 
a(s) = \frac{G(s)}{D(s)}, \label{r1}
\end{equation} 
and ignoring the  contribution to $\Delta a(s)$ from the exchange channels, one finds \mbox{$\Delta G(s)=0$}, {\it i.e.} $G(s)$ is an analytical function in the entire complex plane.  Thus if Eq.(\ref{dis2}) is considered as an integral equation for the KT amplitude, in general, the solution is not unique since it is parametrized by the class of entire functions $G(s)$. The only restriction on $G(s)$ is that it is bounded by $s^{\delta(\infty)/\pi}$  so that the integral in Eq.(\ref{dis2}) converges. However, since in the KT model the high energy behavior is not constrained Eq.(\ref{dis2}) can  be subtracted  arbitrary number of times, in which case there is no restriction on the large-$s$ behavior of $G(s)$. It is therefore more appropriate to regard the dispersion relation, like the one in  Eq.(\ref{dis2}) as a constraint on the amplitude rather than as a dynamical  equation for the amplitude. 

Keeping the exchange terms, with $\Delta a(s)$ given by  the full expression of Eq.(\ref{dis1}) and with the parametrization \mbox{$a(s) = G(s)/D(s)$}, the discontinuity of the function $G(s)$ is given by the exchange terms~\cite{Mandelstam:1960ua,Frazer:1959ua}, 
\begin{align}
\Delta G(s) & = \frac{2\,\rho(s)\,N(s)}{K(s)} \int_{t_-(s)}^{t_+(s)} dt \frac{G(t)}{D(t)} \nonumber \\
& =  \frac{2\,\rho(s)\,f^{*}(s)\,D^{*}(s)}{K(s)} \int_{t_-(s)}^{t_+(s)} dt \frac{G(t)}{D(t)}, \label{gg}
\end{align}
and 
\begin{equation}
G(s) = \frac{1}{\pi} \int_{4m^2}^\infty ds' \frac{\Delta G(s')}{s' - s}.  \label{disg} 
\end{equation} 
The main advantage of using Eq.(\ref{r1}) is that the integral in Eq.(\ref{disg}) depends on the two-body   partial wave amplitude  $f(s)$ in the physical region, {\it i.e.} on the real axis above the elastic threshold where $f(s)$ is entirely determined by the phase shift. The disadvantage is that computations involve a double integral, one in Eq.(\ref{gg}) to obtain the discontinuity, $\Delta G(s)$ and the other in Eq.(\ref{disg}). 

An alternative representation for $a(s)$ is obtained by  writing it in terms of $f(s)$ instead of the denominator, $D(s)$ function, 
\begin{equation} 
a(s) = f(s)\,g(s).  \label{r2} 
\end{equation}
If the left hand cut in $f(s)$ is ignored, {\it i.e.} $N(s)$ as approximated by an entire function then the discontinuity of $g(s)$ becomes proportional to that of $G(s)$ since,  comparing Eq.(\ref{r1}) and Eq.(\ref{r2}) one finds  \mbox{$G(s) = N(s) g(s)$}.  In general, however, $g(s)$ must absorb the cuts of $N(s)$ so that the latter are absent in $a(s)$, which by construction has only the unitary cut. One therefore finds, 
\begin{equation}  
\Delta g(s) = \Delta g_L(s)  +  \Delta g_R(s) ,
\end{equation}
where 
\begin{align} 
\Delta g_R(s) & =   \frac{2\,\rho(s)}{K(s)} \int_{t-(s)}^{t_+(s)}  dt \, g(t) f(t),  \label{gr}\\
\Delta g_L(s)  & =  - \frac{ \Delta N(s)}{N^*(s)}\,g(s)  \label{gl},
\end{align} 
and the dispersion relations for $g(s)$ follows directly from Eqs.(\ref{gr}),(\ref{gl}), 
\begin{align} 
   g(s) &= g_L(s) + g_R(s) \nonumber \\
 &  = \frac{1}{\pi} \int_{-\infty}^{s_{L}} ds' \frac{ \Delta g_L(s')}{s' - s} + 
\frac{1}{\pi} \int_{4m^2}^\infty ds' \frac{ \Delta g_R(s')}{s' - s}. \label{glr}
\end{align}
Here $s_{L}$ marks the beginning of the left hand cut branch point of $f(s)$. At first sight, it  seems that representation given by Eq.(\ref{r2}) is more complicated compared to that of Eq.(\ref{r1}) since the former requires knowledge of the left-hand cut contribution to the two-body scattering amplitude.  Even though precise knowledge of the left-hand cut is necessary for determining the low-energy behavior of the scalar-isoscalar $\pi\pi$ amplitude  and, for example, properties of the  $\sigma$ meson \cite{Caprini:2005zr, GarciaMartin:2011jx}, it does not need to be  the case when the left-hand cut is far away from the physical region. In this case, the left-hand contribution to the physical region could be well approximated by a suitably chosen conformal expansion \cite{Yndurain:2002ud,Gasparyan:2010xz,Danilkin:2011fz}. In the following section, we test the sensitivity of the KT equation solution to the particular form of the left-hand cut. We also remark that Eq.(\ref{r2}) sidesteps computation of the MO function, which would require knowledge of  amplitude phase  at high energies.  Nevertheless, it still requires an analytical representation for the $S$-wave two-body scattering amplitude $f(s)$ in the complex plane, which appears under the integral for $\Delta g_R(s)$  in Eq.(\ref{gr}).

In both cases, either using Eq.(\ref{r1}) or  Eq.(\ref{r2}),  evaluation of $a(s)$ requires computation of a double integral and the equations for $G(s)$ or $g(s)$ can be solved only by iterations. The double integral is of the type 
\begin{equation} 
I(s) = \int_{4m^2}^{\infty} \frac{ds' }{s' - s} \mathcal{A}(s') \int_{t_-(s)}^{t_+(s)} dt \mathcal{B}(t) , \label{doubIntI}
\end{equation} 
where $\mathcal{B}(t)$, which is proportional to the unknown function $G(t)$ or $g(t)$, has a cut on the real axis for  \mbox{$t > 4m^2$}.  The integral over $s'$  runs over the real axis and the $t$ integral runs over the complex contour $\Gamma$ shown in Fig.~\ref{fig:2}. As shown in ~\cite{Pasquier:1968kt,Aitchison:1978ua} the two contours can be deformed simultaneously in such a way that the real $s'$ integration is deformed onto a contour $C'$ in the complex $s'$ plane while the $t'$ complex contour is brought onto the contour $\Gamma'$ along the real axis with the location of $t$ along $\Gamma'$  depending on the location of $s'$ on $C'$. More details of the deformation of contours are summarized in Appendix \ref{kernel}. This procedure allows to interchange  the order of $s'$ and $t$ integrals, and it is referred to as the Pasquier inversion. It results in the following expression of $I(s)$, 
\begin{equation} 
I(s) = \int_{\Gamma'} dt\,\mathcal{B}(t) \int_{C'(t)} ds' \frac{\mathcal{A}(s')}{s' - s}. 
\end{equation} 
The integral over $C'$ is now independent of the function $\mathcal{B}(t)$ and for specific $\mathcal{A}(s')$ the $s'$ integration can be done analytically or numerically resulting in a kernel function $K_\mathcal{A}(s,t)$,  giving 
\begin{equation} 
I(s) =  \int_{\Gamma'} dt\,\mathcal{B}(t)\,K_\mathcal{A}(s,t).  
\end{equation} 
Using this method computation of $G(s)$ or $g(s)$ reduces to a single integral equation which can be solved either by iterations or by matrix inversion. 

The Pasquier inversion seems to favor the representation of Eq.(\ref{r2}) over that given in Eq.(\ref{r1}). This is because the dispersion relation for $g_R(s)$, with  \mbox{$\mathcal{B}(t) =  g(t) f(t)$} and \mbox{$\mathcal{A}(s')  = 2\rho(s')/K(s')$}  results in a 'universal' kernel, $K_g$ {\it i.e.} one that does not depend on the input two-body amplitude, 
\begin{equation} 
K_g(s,t) =2\,\theta(t)\,\Delta_{g}(s,t)  - 2\,\theta(-t)\,\Sigma_{g}(s,t) 
\end{equation}
and the functions $\Delta_{g}$ and $\Sigma_{g}$ are given by Eqs.(\ref{delta}) and (\ref{sigma}) respectively. The integral equation for the function $g(s)$ is then given by, 
\begin{align}
g(s)  =&   -\frac{1}{\pi} \int_{-\infty}^{s_{L}} ds' \frac{1}{s' - s}  \frac{ \Delta N(s')}{N^*(s')}\,g(s')  \nonumber \\
&+ \frac{1}{\pi} \int_{-\infty}^{(M-m)^2} dt\,g(t)\,f(t)\,K_g(s,t).  \label{pasqg}
\end{align} 
On the other hand, the function $\mathcal{A}(s')$ entering the  dispersive representation for $G(s)$, does depend on the input $f(s)$ amplitude, 
\mbox{$\mathcal{A}(s') = 2 f(s')\,\rho(s')\,D(s')/K(s') = 2\,N(s')\,\rho(s')/K(s')$}, leading to 
\begin{align}
& G(s) =\frac{1}{\pi} \int_{- \infty}^{(M-m)^{2}} dt\,\frac{G(t)}{D(t)}\,K_{G}(s,t),    \label{pasqG} 
\end{align} 
where 
\begin{align} 
K_G(s,t) &=2\,\theta(t)\,\Delta_{G}(s,t)  - 2\,\theta(-t)\,\Sigma_{G}(s,t) 
\end{align}
and the functions $\Delta_{G}$ and $\Sigma_{G}$, given by Eqs.(\ref{deltaG}) and (\ref{sigmaG}),  depend on the input two-body amplitude. In this case the kennel $K_G(s,t)$ is model dependent and in general has to be computed numerically.

\subsection{ Model analysis}\label{modelanalysis} 
In the previous section we discussed two common parametrizations of the KT partial wave amplitudes.  These parametrizations are distinguished by how the elastic cut is implemented; either through the MO function, as in Eq.(\ref{r1}),  or using the  partial wave  two body scattering amplitude, as in Eq.(\ref{r2}). It is also possible to compute $a(s)$ directly from Eq.(\ref{dis1}) and (\ref{dis2}), {\it i.e.} without factoring out the elastic scattering amplitude contribution to the KT amplitude. 
  
Given input  partial wave amplitude, $f(s)$ and boundary conditions ({\it cf.} see discussion in Sec.~\ref{methods}) the solution is obtained by numerically solving an integral equation.    The method based on the  Pasquier inversion is potentially best suited for analysis of large data sets, since it reduces  the problem to a one-dimensional integral equations that can be solved by matrix inversion. Furthermore for this method the representation of Eq.(\ref{r2}) is most natural since the kernel functions are universal in this case. However, the input involves p.w. two-body amplitude outside the physical region. The amplitude in the physical region contributes only to the second  integral on the {\it r.h.s} of Eq.(\ref{pasqg})  in the interval between $4m^2$ and $(M-m)^2$. The solution based on Eq.(\ref{pasqG}) also depends on the extrapolation outside the physical region. The integral over $t$ involves the MO function below threshold and the kernel function $K_G(s,t)$ depends on the two-body amplitude evaluated along the complex contour $C'$ ({\it cf.} Eqs.(\ref{deltaG}) and (\ref{sigmaG})). Since the extrapolation of the two-body amplitude is largely-model dependent it  is of interest to study sensitivity of the various representations discussed above to models of the left-hand cut that determines the input amplitude $f(s)$ outside the physical region. 
    
In this section we use a simple analytical model  for $f(s)$ to analyze this sensitivity.  The model  amplitude we use  is dominated by a single resonance and incorporates the left hand cut in an analytical form. 
\begin{figure}
\includegraphics[width=0.54\textwidth]{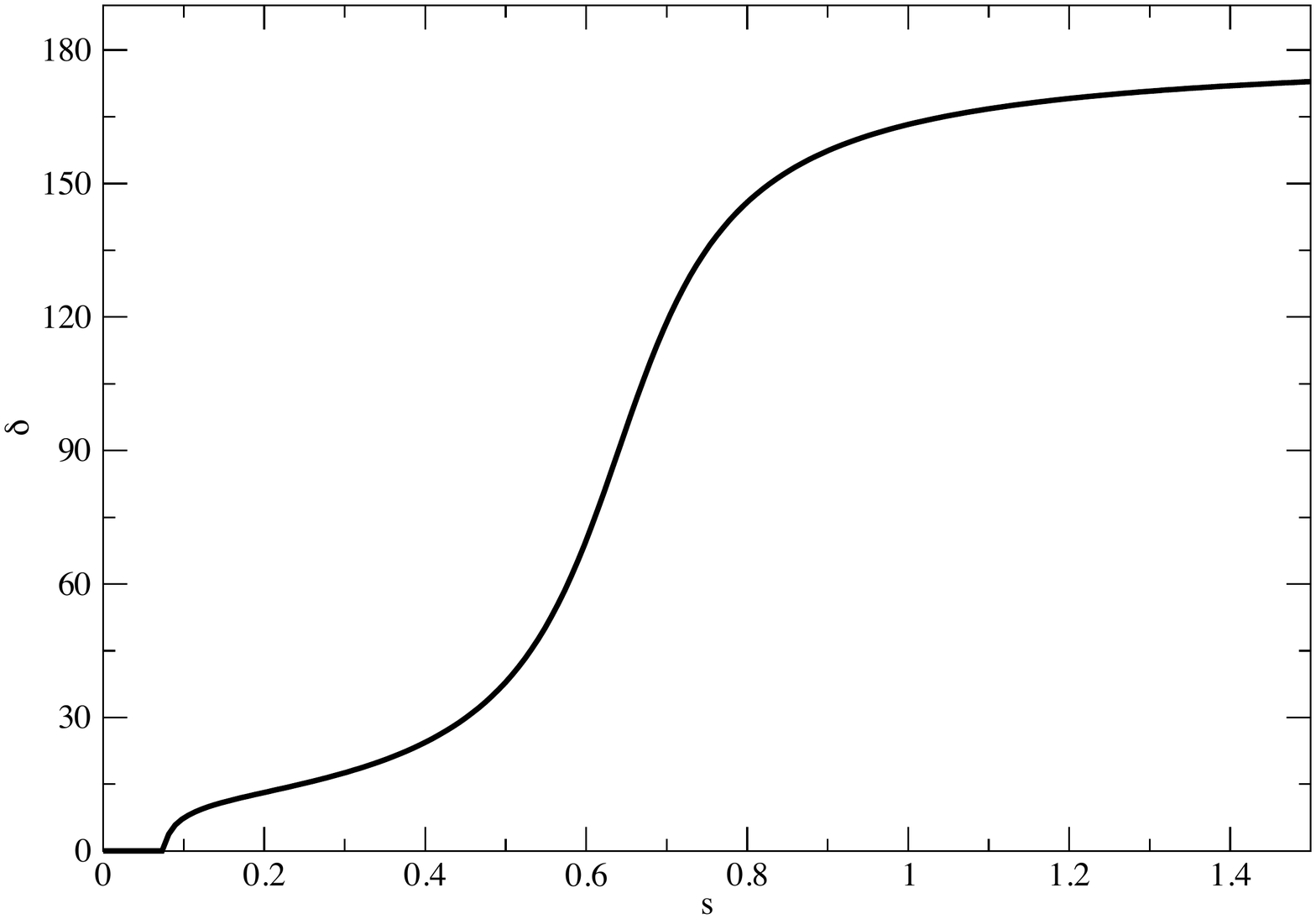}
\includegraphics[width=0.54\textwidth]{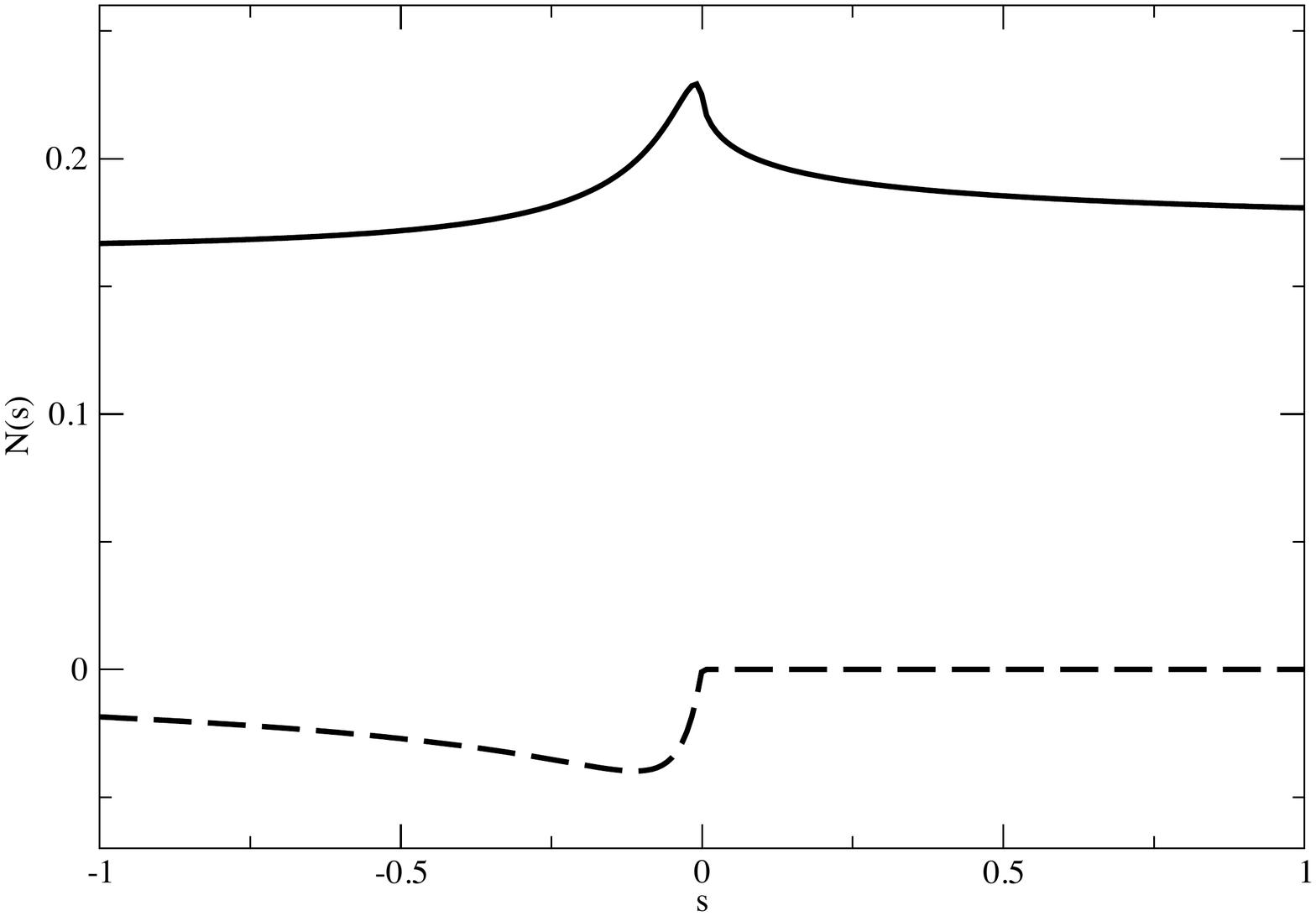}
\includegraphics[width=0.54\textwidth]{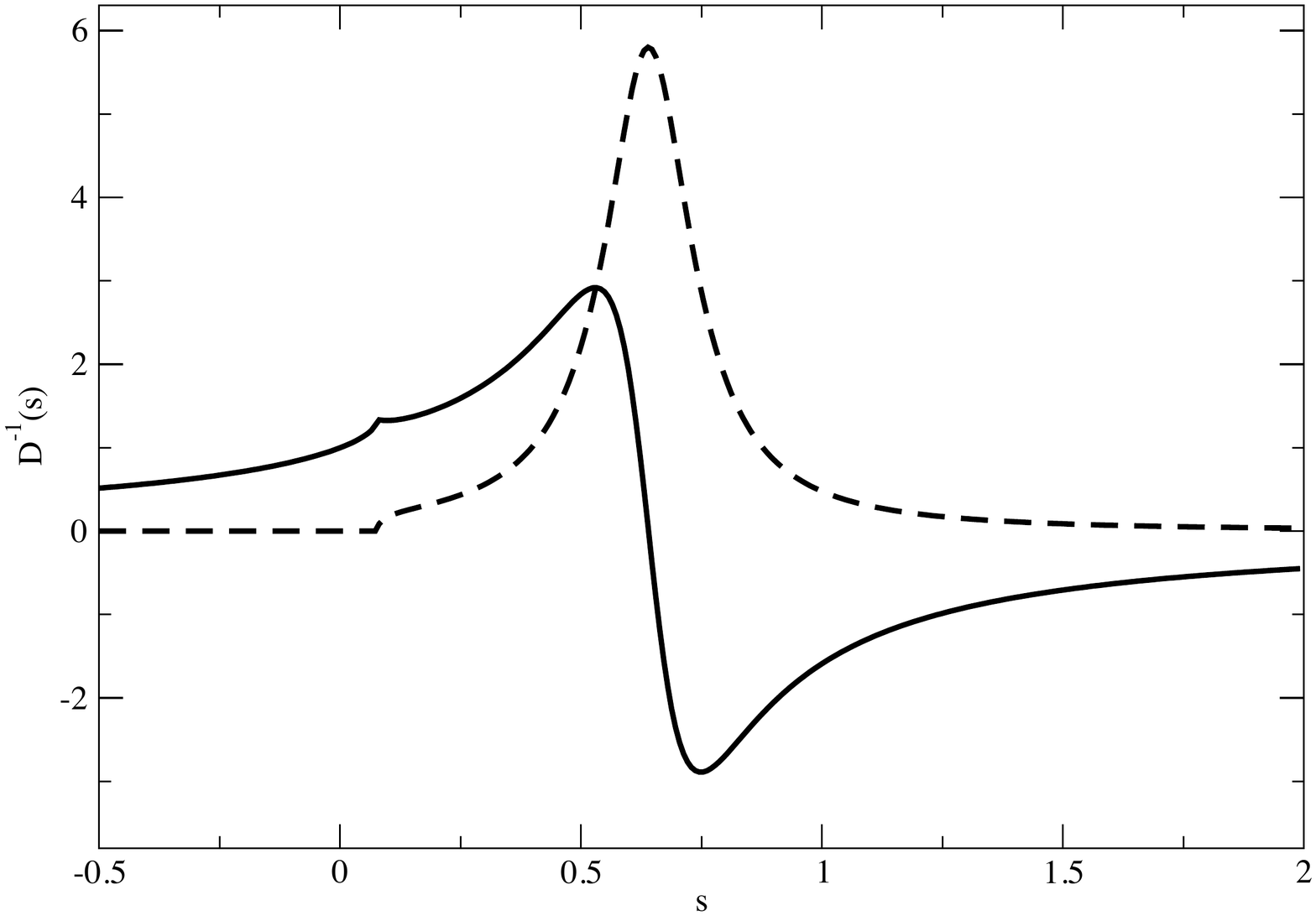}
\caption{Phase shift (top pane), real part (solid curve) and imaginary part  (dashed curve) of  \mbox{$N(s)$} (middle pane) and $1/D(s)$ (bottom pane) corresponding to the model amplitude of Eq.(\ref{t0model}).
\label{NandD}}
\end{figure}
\begin{figure}
\includegraphics[width=0.54\textwidth]{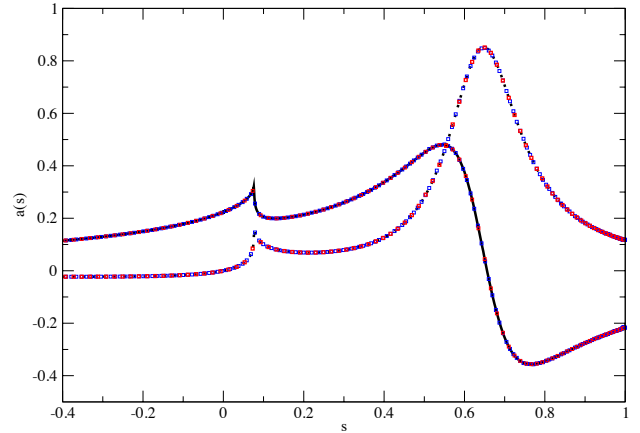}
\caption{Real part (solid) and imaginary part  (dashed) of the function 
\mbox{$a(s)$} obtained by solving Eq.(\ref{pasqg}). Also shown are the solutions, red and blue squares,  obtained by solving directly the integral Eq.(\ref{dis2}) and  Eq.(\ref{pasqG}) respectively.  
\label{fig:sol}}
\end{figure}
\begin{figure}
\includegraphics[width=0.54\textwidth]{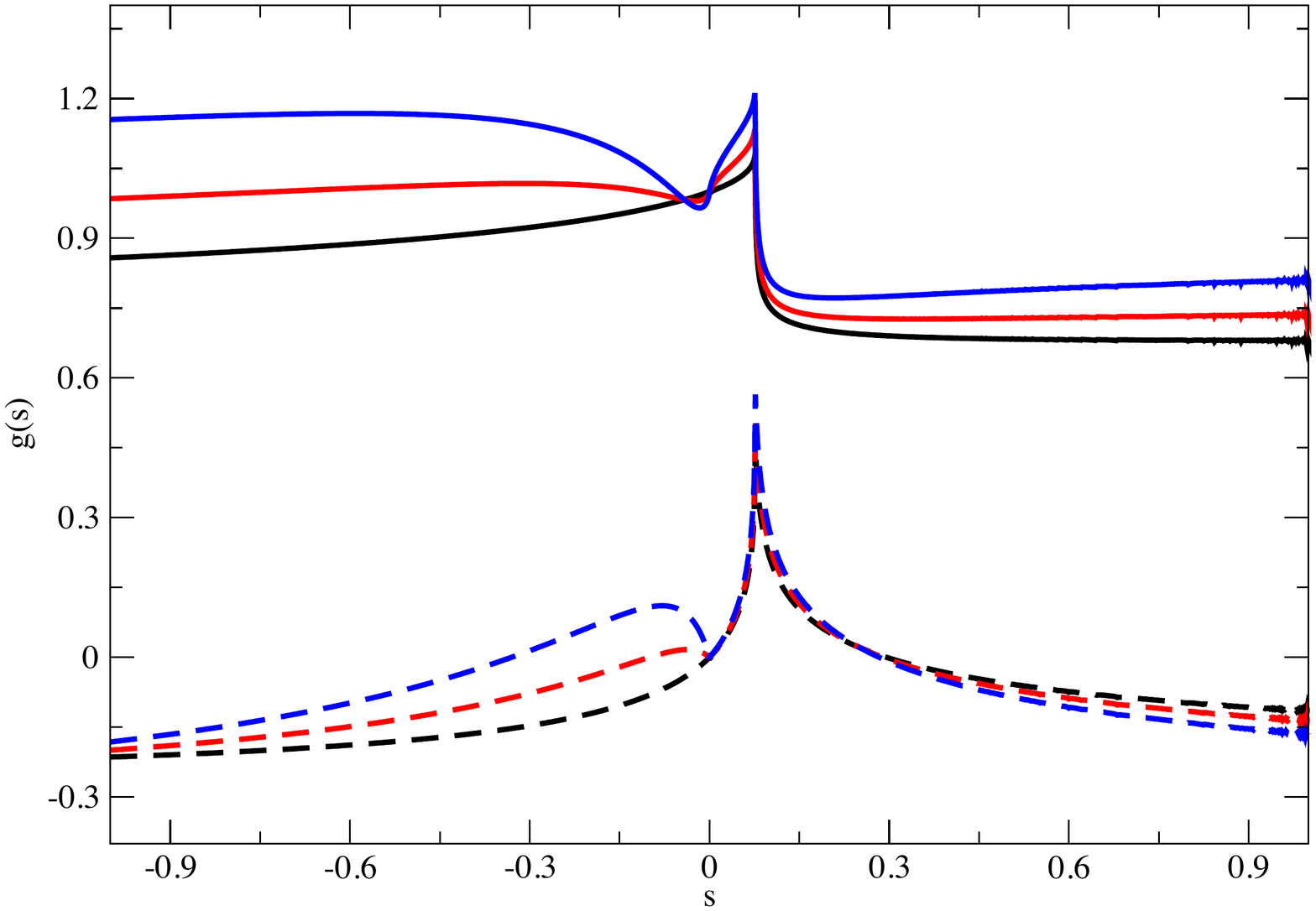}
\includegraphics[width=0.54\textwidth]{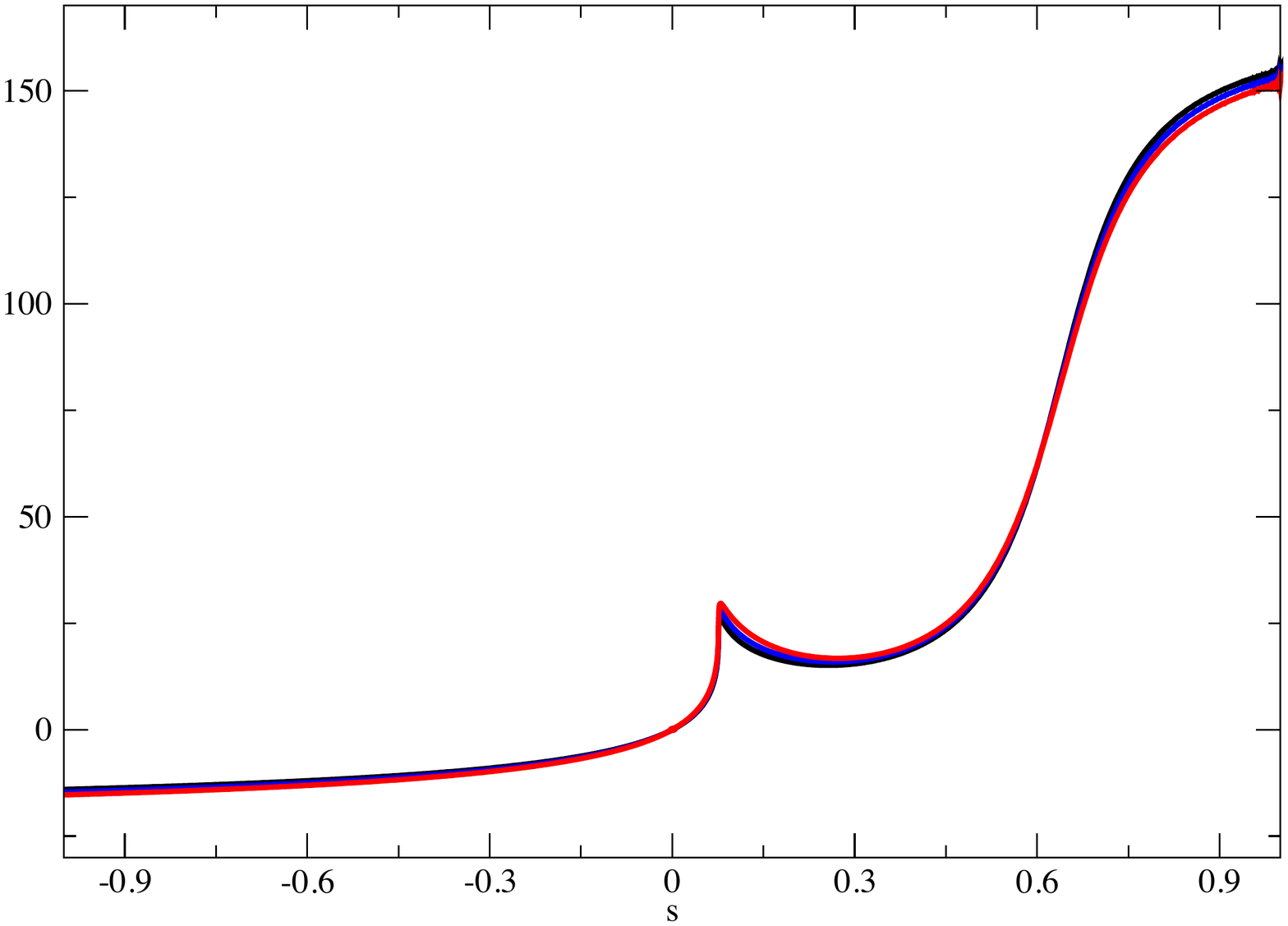}
\caption{Real part (solid) and imaginary part  (dashed) of  the functions,  $g(s)=a(s)/f(s)$  with $\beta$ varied in the range $[0,0.2]$. The colored curves correspond to  \mbox{$\beta = 0$}\,(black), $0.1$\,(red), $0.2$\,(blue).  The phase of $a(s)$ (lower pane), \mbox{$\phi(s) = \tan^{-1} ( \mbox{Im} a(s)/ \mbox{Re} a(s))$},  the color scheme is same as in upper panel.
\label{fig:Nsol}}
\end{figure}
\begin{figure}
\includegraphics[width=0.54\textwidth]{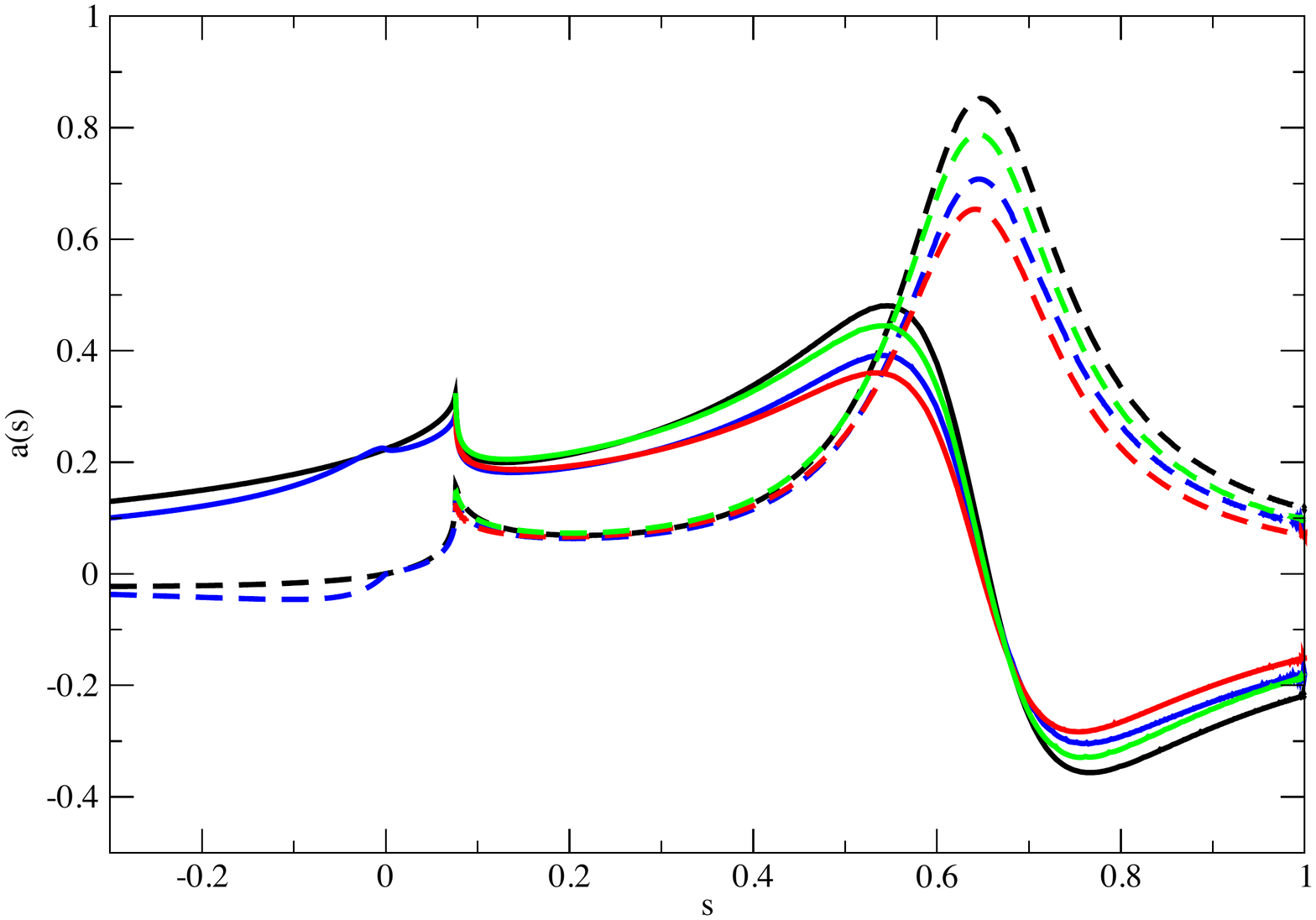}
\includegraphics[width=0.54\textwidth]{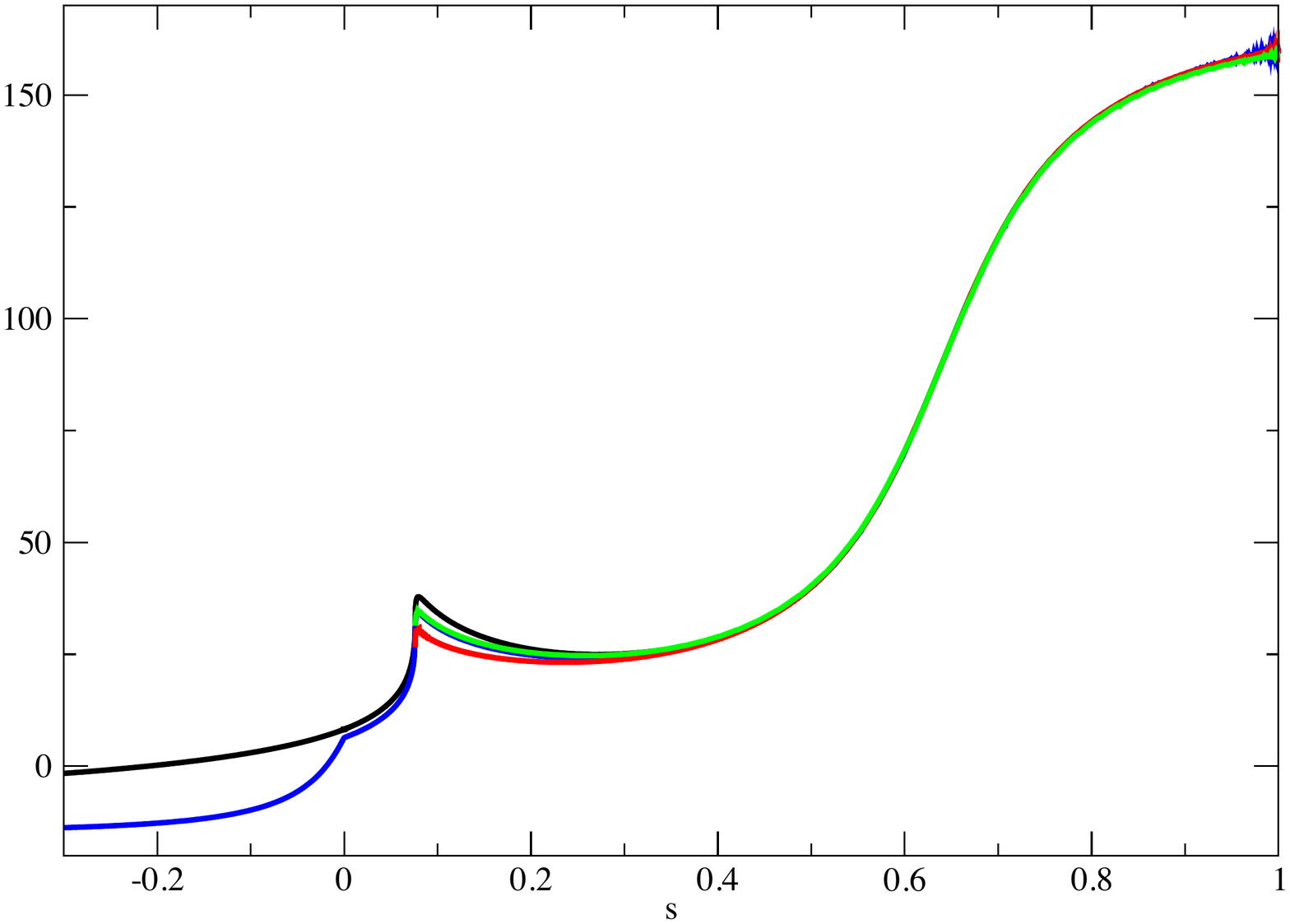}
\caption{   Real part (solid) and imaginary part  (dashed) of the function 
\mbox{$a(s)$} (upper pane) with different approximate schemes for  \mbox{$\beta=0.2$}, black, blue, red and green curves represent full solution,  solution of Eq.(\ref{noleftg}),   solution of Eq.(\ref{cutoffg}) and the solution of Eq.(\ref{cutoffG}) respectively.  The phase of $a(s)$ (lower pane), \mbox{$\phi(s) = \tan^{-1} ( \mbox{Im} a(s)/ \mbox{Re} a(s))$}, the phases are shifted by a constant, so that $\phi(m_{R}^{2})=\pi$, the color scheme is same as in upper pane. 
\label{fig:approxsol}}
\end{figure}
Specifically we choose, $f(s)$ to be given by, 
\begin{align}\label{t0model} 
&f(s) =  \frac{ \alpha }{m_{R}^{2} -s -    \alpha\, \Delta  I_{R}(s)  -      \beta\,\Delta I_{L}(s) } , 
\end{align}
where $\Delta I_{R,L}(s) = I_{R,L}(s) -  \mbox{Re} I_{R,L}(m_{R}^{2}) $, so that the real part of the denominator vanishes at the resonance mass, $s=m^{2}_R$. The function $I_{R}$ is given by the integral over the two-body phase  space 
\begin{equation} 
I_{R}(s) = \frac{s}{\pi} \int_{4 m^{2}}^{\infty} d s' \frac{\sqrt{1- 4 m^{2}/s'}}{s'(s'-s)}. 
\end{equation} 
$I_R(s)$ is responsible for the right hand cut and guarantees unitarity. The contribution from the left hand cut is modeled after a Born-term exchange contribution to the inverse amplitude with a threshold at $4m^2$ in the crossed channel, which results in the left hand cut in the direct channel partial waves starting at $s_L=0$ 
\begin{equation} 
I_{L}(s )=  \frac{1}{\pi} \frac{s}{s-4 m^{2}} \ln \frac{4 m^{2}}{s} .
\end{equation} 
For large-$s$, $f(s) = O(1/s)$  and the phase shift resulting from this model, \mbox{$\delta  = \tan^{-1} (\mbox{Im} f / \mbox{Re} f)$}  approaches $\pi$. 
For a particular choice of parameters,  \mbox{$\alpha=0.1$}, \mbox{$\beta=0.2$}, \mbox{$m_{R}=0.8 \mbox{ GeV}$}, 
and \mbox{$m=0.14\mbox{ GeV}$} and \mbox{$M=1.14\mbox{ GeV}$}, the phase shift,  and the functions $N(s)$ and $D^{-1}(s)$  are shown in Fig.~\ref{NandD}.

We numerically solve Eqs.(\ref{dis2}), (\ref{pasqg}) and (\ref{pasqG}). The boundary condition on the solution is imposed by demanding \mbox{$a(0) = f(0)$} which we  obtain by subtracting the dispersion relations at $s=0$.  As expected, regardless of representation, the three methods give the same result as illustrated in Fig.~\ref{fig:sol}. 
 
We study the sensitivity to the left hand cut by first analyzing its effect on the exact solution. We do this by varying the parameter $\beta$ in Eq.(\ref{t0model}), which controls its strength, {\it i.e.} $\beta = 0$ results in an amplitude $f(s)$ without left hand singularities. This dependence is illustrated in Fig.~\ref{fig:Nsol}. Even though the left-hand cut contribution has some effect on $g(s)=a(s)/f(s)$ in physical decay region, which is  as large as about $15\%$ the phase of $a(s)$ is barely affected by left-hand cut of $f(s)$ (only about $1\%$ change), as seen in lower plot in Fig.~\ref{fig:Nsol}.  
 
In the following  we show how the solutions from different representation of the KT equations change depending on  left-hand cut approximations. Since the KT amplitude can be represented in terms of either $D(s)$ or $f(s)$ we expect a difference in the KT amplitude $a(s)$ depending on how the left-hand cut is approximating in the two representations. In the extreme case, which we study here, in solving the dispersion relation for $G(s)$ or $g(s)$ we restrict the integrals to be as much as possible determined by the physical region of the two-body amplitude. That is, instead of Eq.(\ref{pasqg}) we first use 
\begin{align}\label{noleftg}
g(s)  =&   \frac{1}{\pi} \int_{-\infty}^{(M-m)^2} dt \, g(t) f(t) K_g(s,t).
\end{align} 
where, as a simple approximation, we parametrized contribution from $\Delta g_L$ by a constant.  In principle this constant should be fitted to the data. However, here we study the extreme case when subtraction constant is always fixed by $a(0)=f(0)$ that corresponds to a scenario when the contribution from $\Delta g_L$ is set to be zero. In second study we further limit the range of integration to only include the physical region, 
\begin{align}\label{cutoffg}
g(s)  =&    \frac{1}{\pi} \int_{4 m^{2}}^{(M-m)^2} dt \, g(t) f(t) K_g(s,t). 
\end{align} 
As far as studying sensitivity of the representation in terms of $D(s)$ to the unphysical region in the two-body amplitude in Eq.(\ref{pasqG}) we restrict the integration range to the physical region as well, 
 \begin{align}\label{cutoffG}
& G(s) =\frac{1}{\pi} \int_{4 m^{2}}^{(M-m)^{2}} dt\, \frac{G(t)}{D(t)} K_{G}(s,t).
\end{align} 
The results are shown in Fig.~\ref{fig:approxsol}. The overall shape of the solutions seems to be quite insensitive to the left-hand cut, also see the phase of   solutions  in lower plot in Fig.~\ref{fig:approxsol}. The normalization can vary by as much as  $10-25\%$, however, this can be reduced by changing the subtraction constant. Indeed, the subtraction constant $a(0)$ may be computed approximatively in perturbation theory (such as $\chi$PT) or may be adjusted to empirical data directly. This is an important result as it shows that unitarity in the physical region, where it can be constrained by the data, plays a key role in determining the solutions of the KT equations.

\begin{figure}
\includegraphics[width=0.54\textwidth]{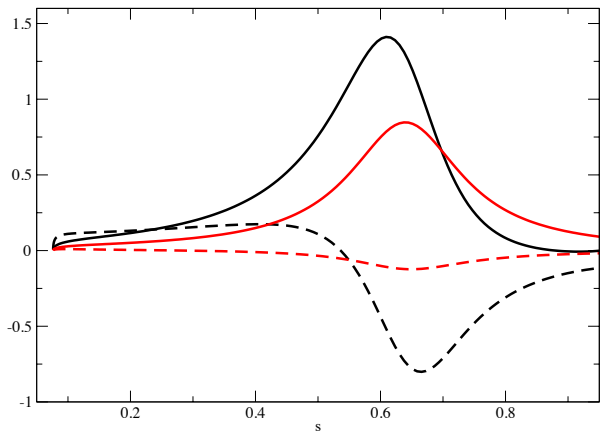}
\caption{  The comparison of real part (solid) and imaginary part (dashed curves ) of $\Delta a(s)$ (black) given by Eq.(\ref{dis1}), and $\rho(s) f^{*}(s) a(s)$ (red). $a(s)$ is the solution of  Eq.(\ref{dis2}) with \mbox{$\beta=0.2$}.
\label{discontF}}
\end{figure}

\begin{figure}
\includegraphics[width=0.54\textwidth]{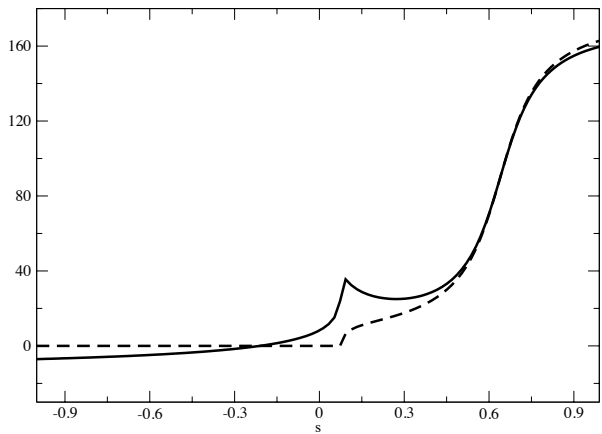}
\caption{  The comparison of phase of $f(s)$, \mbox{$\phi=\tan^{-1}  (\mbox{Im} f/ \mbox{Re} f)$} (dashed black), and phase of $a(s)$ (solid black) by setting \mbox{$\beta=0.2$}. The $\phi$ is normalized to \mbox{$\phi(m_{R}^{2})=\pi$} for comparison purpose.
\label{phaseF}}
\end{figure}

 \begin{figure}
\includegraphics[width=0.47\textwidth]{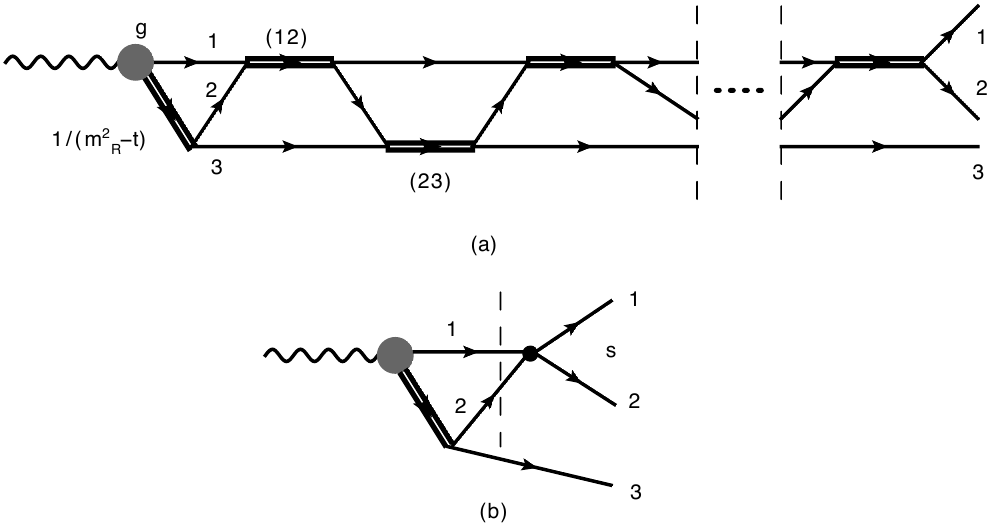}
\caption{   (a) A diagrammatic representation of Eq.(\ref{triangleg}), (b) A triangle diagram with a fixed internal mass $m_{R}$ in $(23)$ sub-channel, this diagram may be associated to the function  \mbox{$ \int_{4 m^{2}}^{\infty} d s'  \frac{1 }{s'-s}\frac{\rho(s')}{K(s')} \int_{t_{-}(s')}^{t_{+}(s')}  d t     \frac{1}{m_{R}^{2}-t}     $}.
\label{sumtriangle}}
\end{figure}

 \begin{figure}
\includegraphics[width=0.54\textwidth]{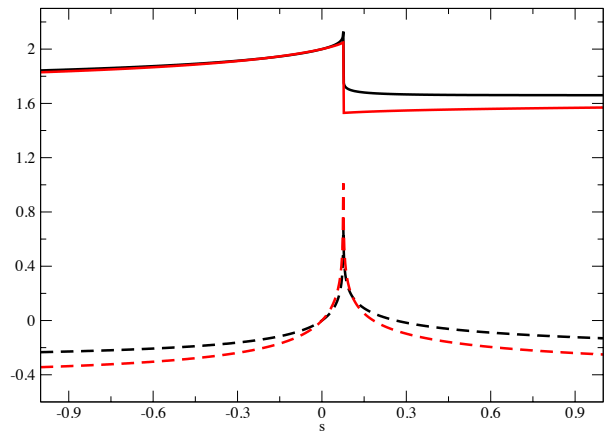}
\caption{  The comparison of real part (black solid curve) and imaginary part (black dashed curve) of $g_{\mbox{pole}} $ given by Eq.(\ref{triangleg}), and the triangle diagram (red curves), with $\alpha=0.1$ and $m_{R} =0.8$GeV. The real part of solutions are shifted up by a unit for the better displaying purpose.
\label{G0}}
\end{figure}

\section{Watson's theorem in a three-body final state}\label{watson}
  
Amplitudes involving two particles in the final states satisfy Watson's theorem, which states that the phase of the corresponding partial wave amplitude in the elastic region coincides with the elastic phase shift \cite{Watson:1952wt}. This is a straightforward consequence of unitarity.  The phase relation does not hold, however, even in the elastic region, when more than two particles are produced and there are interactions among all produced particles~\cite{Aitchison:1965wt,Aitchison:1967wt,Amado:1967wt,Aitchison:1968wt}. As  a simple illustration, one can consider three particle production in a two-body collision: \mbox{$a+b\to c+d+e$}. Partial waves in the $cd$ two-particle subsystem are labeled by the total spin, $J$ and helicity $\Lambda$. They depend on the total center of mass energy squared, $s_{cd}$ of the $cd$ subsystem, momentum transfer $t_{be}$ between particles $b$ and $e$ and total invariant mass squared, $s_{ab}$. In the physical limit, of positive, and say large-$s_{ab}$ and small-$s_{cd}$, {\it i.e.} in the $cd$ elastic region, Watson theorem would relate the phase of the $cd$ partial wave amplitude  with spin $J$ to that of the phase shift of the spin-$J$, $cd$ elastic scattering amplitude. However, since $s_{ab}$ is also in the physical region,  phase of the amplitude also depends on contributions from discontinuities in the $ab$ channel, {\it i.e.} all intermediate processes  \mbox{$a+b \to X  \to c + d +e $}, which couple at a given value of $s_{ab}$. In particular, in the limit of large-$s_{ab}$ this phase can be determined from Regge-theory~\cite{Brower:1974rt}. The nonequivalence between imaginary part of the amplitude and its discontinuity was observed early on when unitary relations for multiparticle amplitudes were investigated, for example in ~\cite{Blankenbecler:1961us,Cook:1962us}.  For the  recent discussions we refer to \cite{Schneider:2012ez, Ananthanarayan:2014pta,Caprini:2015wja}.
 
There is significant interest in using dispersion relations to extract two-particle phase-shifts from amplitude analysis of three-particle final states, {\it e.g.} produced in pion or photon diffraction or in heavy flavor decays~\cite{Brian:2006wt,Cleoc:2008wt,Caprini:2006wt,Link:2007wt,Magalhaes:2011wt,Hass:2011wt,Uhl:2014wt}.

In general, three-body  rescattering  effects between resonance decay products and spectators modify the discontinuity relation that follows from two-body unitarity and as a consequence lead to violation of Watson's theorem.  We can use the specific model described in the previous section study to investigate the size of these effects. Using the solution from Eq.(\ref{dis2}) with $\beta=0.2$, in Fig.~\ref{discontF}, we compare $\Delta a(s)$ given by Eq.(\ref{dis1}) and $\rho(s) f^{*}(s) a(s)$. As can be clearly seen,  $\Delta a(s)$ does have an imaginary part. It arises from the $t$ and $u$-channel exchanges, {\it i.e.} rescattering between a pion emerging from a decay of a $t$ or $u$ channel resonance an the spectator.   
 
We also plot the phase of $a(s)$ compared to the phase of $f(s)$ in Fig.~(\ref{phaseF}). The phase difference of $a(s)$ and $f(s)$ is given by the phase of \mbox{$g(s)=a(s)/f(s)$}. Above the elastic threshold, the phase difference is due to the cross-channel exchanges. {\it c.f.} Eq.~\ref{dis1}.   Near the resonance, the phase difference between $a(s)$ and $f(s)$ is minor, which is consistent with the expectation that three-body effects are at most a correction when two particles emerge from a resonance. Since $\Delta a(s)\ne 0$ we  observe that $a(s)$ acquires a non-zero phase below the physical threshold $4\,m^{2}$.

The non-vanishing phase of $a(s)$ can be understood by examining the asymptotic behavior of the reduced amplitude $g(s)$.  Even in the case when only the right-hand cut of $f(s)$ is included, as \mbox{$s \rightarrow -\infty$}, we obtain,
\begin{align} 
g(s)   \stackrel{s \rightarrow -\infty}{\longrightarrow}  &- \frac{2}{\pi s} \int_{4 m^{2}}^{\infty} d s'  \frac{
 \rho(s') }{ K(s')} \int_{t_{-}(s')}^{t_{+}(s')}  dt\,f(t)\,g(t).
\end{align} 
In general,  $f(s)$ is a complex function, thus, the phase of $g(s)$ cannot be zero below $4 m^{2}$ as is seen in Fig.~\ref{phaseF}. The cusp at threshold originates from the triangle  (anomalous) singularity ~\cite{Landshoff:1962ct,Eden:1966ct,Wu:2012prl,Wang:2013ct,Wang:2013prl} and it is caused by the reduced amplitude $g(s)$, which can be seen in both Fig.~\ref{fig:sol} and Fig.~\ref{fig:Nsol}.  To examine  this point, in Eq.(\ref{r2}), we replace $f(s)$ by  a simplified expression, $\alpha/(m_{R}^{2}-s )$ and obtain, 
\begin{align}\label{triangleg}
 g_{\mbox{pole}}(s)  & =  \frac{1}{\pi} \int_{4 m^{2}}^{\infty} d s' \frac{1}{s'-s } \nonumber \\
& \times \frac{2\,\rho(s') }{ K(s')} \int_{t_{-}(s')}^{t_{+}(s')}  dt\,\frac{ \alpha }{m_{R}^{2} - t }\,g_{\mbox{pole}}(t)  .
\end{align} 
$g_{\mbox{pole}}(s)$ may be identified with the re-scattering contribution from a perturbative analysis of the triangle diagram as shown in Fig.~\ref{sumtriangle}. In perturbation theory, we approximate   \mbox{$g_{\mbox{pole}}(t)=1$} under the integral on the right-hand side of Eq.(\ref{triangleg}) and this leads to the triangle diagram of Fig.~\ref{sumtriangle}(b). When \mbox{$m_{R}> 4 \,m^2$}  the contour integration  $\Gamma(t)$ will sweep through the real axis and pick up an absorptive part resulting in imaginary part of $g_{\mbox{pole}}(s)$ below threshold \mbox{$s<4\,m^{2}$}  generating  a cusp near threshold.  In Fig.~\ref{G0}, we show the  solution of Eq.(\ref{triangleg}), $ g_{\mbox{pole}}(s) $ and compare it to the first rescattering correction (a triangle diagram).

\section{Summary} \label{summary}
   
We analyzed different representations of the Khuri-Treiman equation and demonstrated, 
on the basis of a simple model, that solutions from various representations are indeed identical. The main focus of our study was the Pasquier inversion, which until recently has never been used in the numerical analysis of the three particle decays. Therefore, we studied the sensitivity of the various representations to the left-hand cut contribution. 
We showed that the Pasquier inversion might be indeed a good complementary approach to the  direct  solution of the KT equations, especially when the physical region does not depend strongly on the accurate form of the left-hand cut.
In the end, we discussed the Watson's theorem when the interaction among  three particles in the final state is involved.  We concluded that the phase of the solution of KT equation did not coincide with the elastic phase shift of scattering amplitude.

\section{ACKNOWLEDGMENTS}
We thank Michael~R.~Pennington for many useful discussions, we also thank Andrew~W.~Jackura for the final proof reading of this manuscript. 
This research was supported in part by the U.S.\ Department of Energy under Grant No.~DE-FG0287ER40365, the Indiana University Collaborative Research Grant and U.S.\ National Science Foundation under grant PHY-1205019.  We also acknowledge support from U.S. Department of Energy contract DE-AC05-06OR23177, under which Jefferson Science Associates, LLC, manages and operates Jefferson Laboratory. 
 
\appendix

\section{Kernel functions}\label{kernel}

\begin{figure}
\includegraphics[width=0.48\textwidth]{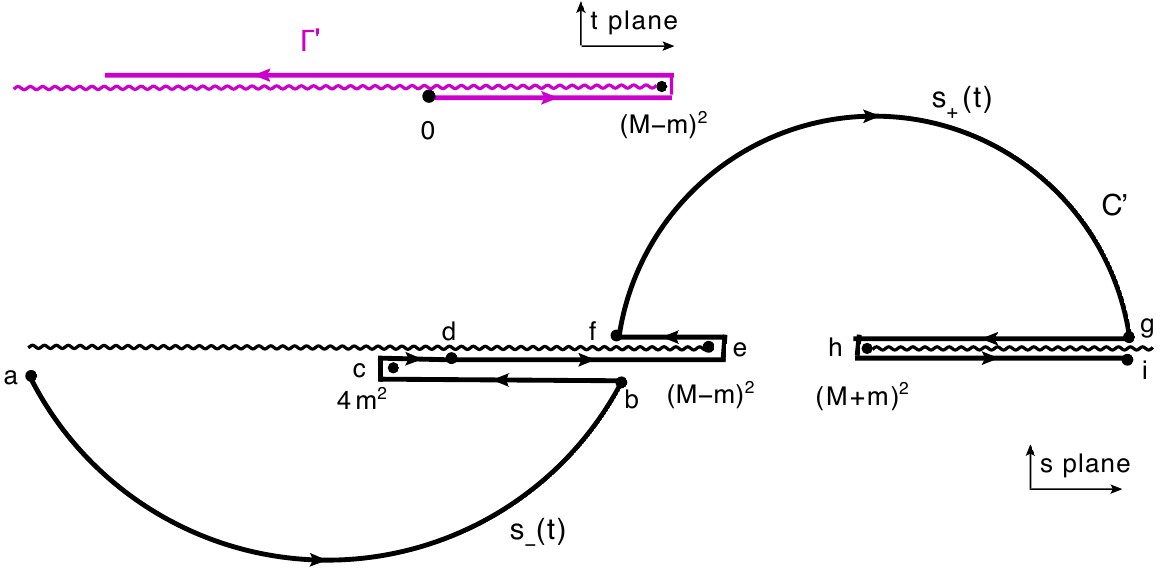}
\caption{The contour $\Gamma'$ (solid purple lines) and $C'$ (solid black curves) and  in Eq.(\ref{collapseintef}). 
 The arrows indicate the directions  that   invariants follow along the path of integrations.  The purple and black wiggle lines represent unitarity cut in $t$-plane and  cuts attached to two branch points: $(M\pm m)^{2}$ in $s$-plane  respectively. The  points labeled by \mbox{$a-f$} correspond to (a) \mbox{$s_{-}(0)= -\infty$}, (b) \mbox{$s_{-}(4\,m^{2})= \frac{M^{2} -m^{2}}{2}$}, (c) \mbox{$s_{-}(\frac{ M^{2} -m^{2} }{2})= 4 m^{2}  $}, (d) \mbox{$s_{\pm}((M-m)^{2})= m\,(m+M) $}, (e) \mbox{$s_{+}(m\,(M+m))= (M-m)^{2} $}, (f) \mbox{$s_{+}(4\,m^{2})= \frac{ M^{2} -m^{2} }{2}$},  (g) \mbox{$s_{+}(0)= \infty$}, (h) \mbox{$s_{+}(m\,(m-M))= (M+m)^{2}$},  and (i) $s_{+}(-\infty)= \infty$, respectively.
\label{collapsepath}}
\end{figure}

In a  double integral equation of the type given by  Eq.(\ref{dis1}), 
\begin{equation}\label{doubleintef}
 I =  \int_{ 4 m^{2}}^{\infty} d s' \frac{1}{s'-s} \frac{N(s') }{U(s')}   \int_{t_{-}(s')}^{t_{+}(s')}  \!\!\!\!\!\!\!\!\!\!\! (\Gamma) \    d t\,a(t),
\end{equation}
the contour $\Gamma$ followed by $t$ integration  is defined in Fig.\ref{fig:2} and the integration path over $s'$ is defined on the real axis.   The square root function $U(z)$ is given by  \mbox{$U(z)=\sqrt{ \left(z - (M-m)^{2} \right)  \left(z -(M+m)^{2} \right)}$}   in the complex-$z$  plane. The phase convention for $U(z)$ is chosen by \mbox{$U(s\pm  i0 ) = (\mp , i , \pm)\,|U(s)|$} for \mbox{$s\in ([-\infty, (M-m)^{2} ]$}, \mbox{$[ (M-m)^{2} , (M+m)^{2} ],\,[ (M+m)^{2} , \infty])$} respectively.  In Eq.(\ref{doubleintef}), the phase of $U(s)$  is chosen as  the value of $U(s)$   right below the two cuts attached to branch points $(M\pm m)^{2}$, {\it i.e.} $K(s)/\rho(s) =U(s-i0) $.

As described in  \cite{Pasquier:1968kt,Aitchison:1978ua}, when the order of the $s'$ and $t$ integrals is reversed, Eq.(\ref{doubleintef}) becomes, 
\begin{align}\label{collapseintef} 
 &  I = \int_{\Gamma'}  dt\,a(t)     \left [  \int_{s_{\Gamma'} (t)}^{\infty}  \!\!  \!\!\!\!\!\!\!\! (C') \ d s'     \frac{1 }{s'-s} \frac{N(s')}{U(s')} \right ] .
\end{align}
The contours $C'$  and $\Gamma'$ avoid the singularities in the integrand, see Fig.~\ref{collapsepath}.  
Whether $s_{\Gamma'} (t)  $ is $s_{+} (t) $ or $s_{-} (t) $ depends on whether $t$ is above or below the cut in $t$-plane respectively,  and $s_{\pm}(t)$ are  given by the solution of \mbox{$\phi (s_{\pm},t)=0$},  where \mbox{$\phi(s,t) = s\,t\,u - m^{2} (M^{2}-m^{2})^{2}$} and $s+t+u =M^{2}+3\,m^{2} $. Splitting the $s'$ integration path,  \mbox{$   \int_{ \Gamma'}   =  \left [  \int_{ 0 _{-}}^{ (M - m)^{2}_{-} }  -    \int^{ (M - m)^{2}_{+}}_{  0_{+} }  \right ]  +\int_{ 0_{+}}^{  - \infty_{+}}        $} in Eq.(\ref{collapseintef}) (subscript $+/-$ of integration limits denotes the path of integration lying above or below the cut attached to branch point $(M-m)^{2}$ in $t$-plane, see Fig.~\ref{collapsepath}), one obtains, 
\begin{align}\label{pasquierintef} 
& I =  \int_{ -\infty }^{ (M - m)^{2} }   d t\,a(t)       \nonumber \\
& \quad \times  \left [   \theta(t) \int_{s_{-} (t)}^{s_{+} (t)}   \!\!\!\!\!\!\!\!\!\!\!\!  (C')  - \theta(-t)  \int_{s_{+} (t)}^{ \infty}   \!\!\!\!\!\!\!\!\!\! (C')  \  \right ]  d s'  \frac{1 }{s'-s}  \frac{N(s')}{U(s')}  .
\end{align}

The kernel functions $\Delta_{G}   $ and $\Sigma_{G} $ are defined by, 
\begin{align}
\Delta_{G} ( s,t) &= \int_{s_{-}(t)}^{s_{+}(t)}  \!\!\!\!\!\!\!\!\ \!\!\!\!\!\!  (C') \ d s'   \frac{1}{s'-s } \frac{ N(s') }{  U(s')  }  \label{deltaG}, \\
\Sigma_{G}  ( s,t) &= \int_{s_{+} (t)}^{ \infty}  \!\!\!\!\!\!\!\!\ \!\!\!  (C') \  d s'  \frac{1}{s'-s }  \frac{ N(s') }{  U(s')  }   \label{sigmaG} .
\end{align}
and Eq.(\ref{doubleintef}) finally becomes
\begin{align}\label{pasqinvert} 
& I =   \int_{ -\infty }^{ (M - m)^{2} } \!\!\!\!   dt\,a(t)      \left (\theta(t)\,\Delta_{G}(s, t) - \theta(-t)\,\Sigma_{G}(s, t)   \right)\,.
\end{align}

In general, kernel functions $\Delta_{G}   $ and $\Sigma_{G} $ have to be evaluated numerically by contour integration in the  complex plane. In particular for the case \mbox{$N(s)=1$}, the corresponding kernels, which we denote as $\Delta_{g}   $ and $\Sigma_{g} $  can be expressed in terms of elementary functions
\begin{align}\label{delta}
\Delta_{g}  & (s, t)   = \frac{1}{U(s )}   \ln \left| \frac{R  (s, t) + U(s )\,U(t  )}{R  (s, t) - U(s)\,U(t) } \right|   - \theta \left ( \phi  (s, t) \right ) \frac{i\,\pi}{U(s )} , 
\end{align}
and
\begin{align}\label{sigma}
& \Sigma_{g} (s, t) = \frac{1}{U(s )} \nonumber \\
 &   \times  \ln  \frac{ \left(s_{+}(t )-s  \right) \left( s- M^{2} -m^{2} + U(s) \right)  }{ \left( s_{+}(t )-s  \right) (s- M^{2} -m^{2} )  + U^{2} (s) - U(s)\,U(s_{+}(t))}   .
\end{align} 
where 
\begin{align}
R &(s, t)  = - M^{4}  +  (s - m^{2}) (t- m^{2})    +M^{2} (s+t  ).
\end{align}
For real $s$ and $t$ the physical values of $\Delta_{g} $ and $\Sigma_{g} $ correspond to the limit $s + i0$ and $t + i0$.


\end{document}